\documentclass[doublecol,figures]{epl2} 

\usepackage{amsmath}
\usepackage{amsfonts}
\usepackage{amssymb}

\usepackage[breaklinks,colorlinks,linkcolor=blue,citecolor=blue,urlcolor=blue]{hyperref}

\newcommand{\FATP}{F$_1$-ATPase}
\newcommand{\mrd}{\mathrm{d}}
\newcommand{\mre}{\mathrm{e}}
\newcommand{\mri}{\mathrm{i}}
\newcommand{\mean}[1]{\left\langle #1 \right\rangle}
\newcommand{\fe}{f^\mathrm{ex}}
\newcommand{\pstat}{p^\mathrm{s}}
\newcommand{\dx}{\varDelta x}

\title{ Fine-structured large deviations and the fluctuation theorem:
  Molecular motors and beyond}
\shorttitle{Fine-structured large deviations and the fluctuation theorem} 

\author{Patrick Pietzonka, Eva Zimmermann \and Udo Seifert
\thanks{E-mail: \email{useifert@theo2.physik.uni-stuttgart.de}}}
\shortauthor{P. Pietzonka \etal}

\institute{                    
  II. Institut f\"ur Theoretische Physik, Universit\"at Stuttgart - 70550
  Stuttgart, Germany
}


\abstract{By considering sub-exponential contributions in large deviation theory,
we determine the fine structure in the probability distribution of the observable
displacement of a bead coupled to a molecular motor. More generally, for any stochastic motion along a
periodic substrate, this approach reveals a discrete symmetry of this 
distribution for which hidden degrees of freedom lead to a periodic modulation of the slope typically 
associated with the fluctuation theorem. Contrary to previous interpretations of 
experimental data, the mean force exerted by a molecular motor is unrelated to
the long-time asymptotics of this slope and must rather be extracted from its short-time limit.
}

\begin{document}

\maketitle

For non-equilibrium steady states, the fluctuation theorem (FT) expresses a
remarkable symmetry of the distribution function for entropy production
as reviewed in \cite{seif12}.
Molecular motors \cite{juel97,kolo07} are one class of systems for which the
FT has been explored. While early work concentrated on theoretical models
\cite{seif05,andr06c,astu07a,lau07a,laco08,laco09,ge12}, more recently in a
potentially intriguing application, the authors have attempted to derive the 
torque exerted by motors like the \FATP{} from an FT-like
representation of the data for displacements in the long-time limit \cite{haya10,haya12,haya13}. 
Entropy production and displacement, however, are strictly proportional to each other only in
systems with a completely flat energy landscape. Since in any more 
realistic description this proportionality is lost, it is not clear
whether displacements should obey an FT-type symmetry and, if they do,
whether and how the characteristic slope is related to entropy production
or to the torque exerted by a motor.

The crucial role that hidden slow degrees of freedom can play for 
apparent deviations from the FT-symmetry for entropy production has been 
studied in \cite{raha07,pugl10,mehl12,espo12,alta12}. In the experimental analysis of molecular motors, 
typically the motion of the motor is accessible only indirectly through 
the observation of the displacements of a colloidal particle elastically 
linked to some (hidden) degree of freedom of the motor. As a main result,
we will identify a discrete symmetry for the 
probability distribution of displacements leading to a modulated slope in 
an FT-like representation. In fact, this symmetry will hold beyond 
molecular motors for any driven stochastic motion along a spatially 
periodic structure. Entropy production, however, can be inferred from such 
data on displacements only if the motor does not exhibit idle cycles. This 
fact will lead to a simple, experimentally accessible, criterion for the 
presence of such cycles.

We derive these results using large deviation theory as reviewed in \cite{touc09}. 
More specifically, we explore the sub-exponential contributions to the large 
deviation function. While the main part of our Letter thus deals with the 
long-time limit, we show in a brief appendix that the force exerted by a 
motor must rather be inferred from data on displacement in the 
{\sl short-time} limit contrary to what has been attempted in \cite{haya10,haya12}.

\begin{figure}
\centerline{\includegraphics[width=\columnwidth]{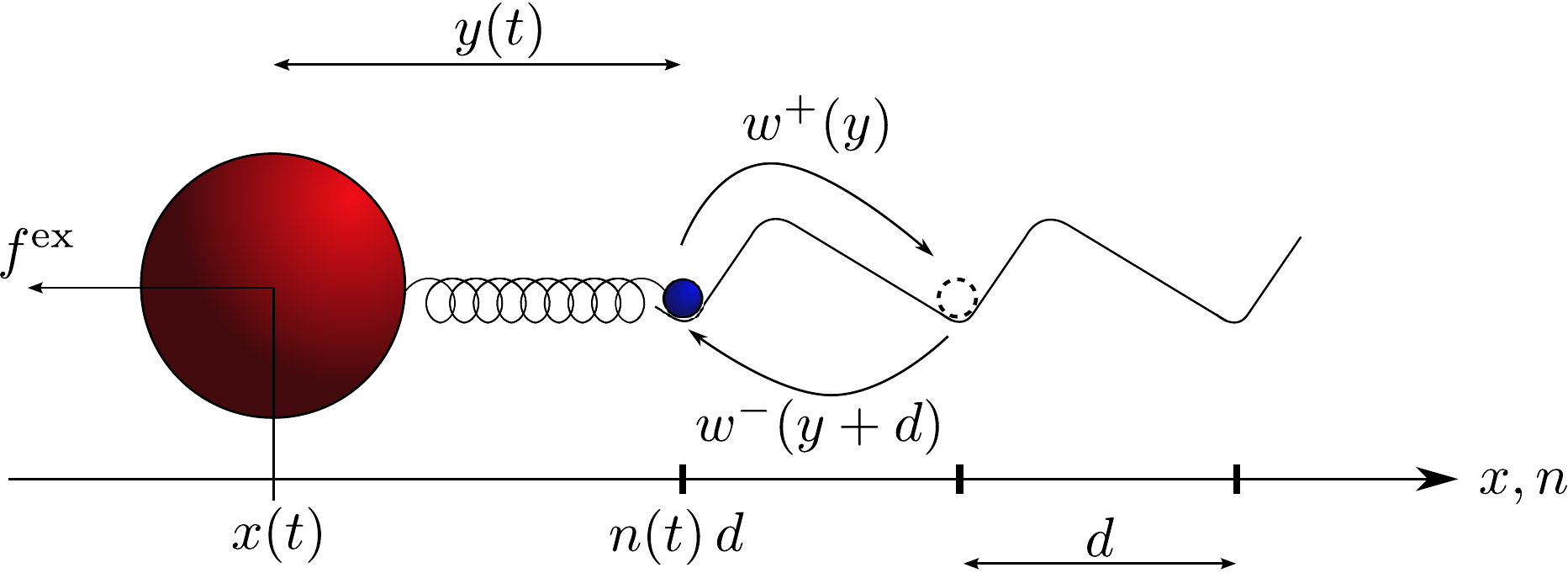}}
\caption{Illustration of the one-step model. A bead (red) at position $x(t)$
  is coupled to
  the motor (blue) at position $n(t)d$. The motor is a distance $y(t)$ ahead of the bead to which an external force $\fe$ is applied. The motor jumps at rates $w^\pm$ between
adjacent minima of its (unresolved) periodic internal potential.}
\label{fig:model}
\end{figure}

The basic idea of our approach is best illustrated using the arguably
simplest model for a molecular motor subject to an external force transmitted
by a probe particle as introduced in \cite{zimm12}, where it was applied to model the measured efficiency of the \FATP \cite{toya10,toya11a}. The generalization to
molecular motors with several substeps \cite{lipo00,liep07a,gasp07,qian07,mull08,astu10,seif11a,golu13} will be given further below. In the
one-step model shown in Fig.~\ref{fig:model}, the time-dependent position of the motor is
given by a discrete variable $n(t)$ whereas the diffusive motion of
the much larger, observable,  bead is described by a continuous position
$x(t)$. Each step of the motor covers the spatial distance $d$,
such that the distance between the motor and the bead is $y(t)=n(t)d-x(t)$.
The diffusive motion of the bead is governed by the Langevin-equation
\begin{equation}
  \partial_t x(t)=[-\fe+\partial_yV(y)+\xi(t)]/\gamma.
  \label{eq:langevin}
\end{equation}
It involves an external force $\fe$, a force
$\partial_yV(y)$ arising from the elastic linker with a potential energy $V(y)$, and thermal noise $\xi(t)$ with the usual
correlations $\langle \xi(t_1)\xi(t_2)\rangle = 2\gamma\delta(t_2-t_1)$ and
$\gamma$ the friction coefficient of the bead. Both Boltzmann's constant $k_\mathrm{B}$ and the temperature
$T$ are set to 1 throughout the letter thus making all energies dimensionless. 
The motor takes discrete forward and backward steps of length $d$
at a rate $w^+(y)$ 
and $w^-(y)$,
respectively. Each forward step involves hydrolyzation of one ATP to
ADP liberating $\varDelta \mu$ of free energy 
and each backward step induces the reversed reaction. Both rates
therefore depend implicitly on the concentrations of these molecules and on the instantaneous
separation $y$. While the specific choice of forward and backward
rate will be irrelevant for our main results, they have to
obey the local detailed balance condition
\begin{equation}
w^+(y)/w^-(y+d)=\exp[\varDelta\mu-V(y+d)+V(y)].
\label{eq:db_motor}
\end{equation}

The evolution of the probability $p(n,y,t)$ then follows the master equation
\begin{equation}
\partial_t p(n,y,t)= (L_1+L_2) p(n,y,t),
\end{equation}
where 
\begin{equation}
L_1\equiv \partial_y[-\fe+\partial_yV(y)]/\gamma+\partial^2_y/\gamma
\end{equation} is the Fokker-Planck operator for motion of the bead and 
\begin{align}
L_2 p(n,y,t)\equiv&+w^+(y-d)\,p(n-1,y-d,t) \nonumber \\
&+w^-(y+d)\,p(n+1,y+d,t)\nonumber\\
&-[w^+(y)+w^-(y)]\,p(n,y,t)
\end{align}
deals with  the discrete jumps of the motor.
As initial condition, we choose $p(n,y,0)=\delta_{n0}\,\delta(y-y_0)$.
Note that the system as a whole, characterized by both $n$ and $y$, does not
reach a genuine non-equilibrium steady state, since we do not introduce
periodic boundary conditions\footnote{For rotary motors, like \FATP{}, $x$ and $n$ correspond to the accumulated number of revolutions.}.  However, the marginal
distribution for the elongation $y$ reaches a stationary distribution
$\pstat(y)$, from which we will later sample the initial value $y_0$.

The long-time limit can be analyzed by introducing the generating
function \cite{vankampen}
\begin{equation}
g(\lambda,y,t)\equiv \sum_{n=-\infty}^\infty \mre^{\lambda n} p(n,y,t),
\end{equation}
which obeys $\partial_tg(\lambda,y,t)= \mathcal{L}(\lambda) g(\lambda,y,t)$ with
\begin{align} 
\mathcal{L}(\lambda) g(\lambda,y,t)
\equiv&+\mre^{\lambda}\,w^+(y-d)\,g(\lambda,y-d,t)\nonumber\\
&+\mre^{-\lambda}\,w^-(y+d)\,g(\lambda,y+d,t)
\nonumber  \\ 
&+[L_1-w^+(y)-w^-(y)]\,g(\lambda,y,t)
\end{align}
with initial condition $g(\lambda,y,0)=\delta(y-y_0)$.

The operator $\mathcal{L}(\lambda)$ has right eigenfunctions $q_i(\lambda,y)$ 
 with eigenvalues $\alpha_i(\lambda)$. In the long-time limit, the largest
eigenvalue (with label $i=0$) will dominate leading to
\begin{equation}
g(\lambda,y,t) \approx  \mre^{\alpha_0(\lambda) t} c_0(\lambda,y_0)\,q_0(\lambda,y)\equiv \mre^{\alpha_0(\lambda) t} Q(\lambda,y,y_0) .
\end{equation}
The last definition combines the expansion coefficient $c_0$
through which the initial condition $y_0$  enters with the dominant eigenfunction
$q_0$. Since $g(0,y,t)$ is the marginal distribution for $y$, we can identify
$Q(0,y,y_0)=\pstat(y)$, which is, in fact, independent of $y_0$. From the generating function, the probability
$p(n,y,t)$ can be recovered using the residue theorem in the complex
$z\equiv\mre^\lambda$ plane as
\begin{align}
 p(n,y,t)&=\frac{1}{2\pi\mri}\oint \mrd z\,z^{-n-1}g(\ln z,y,t) \nonumber\\
&\approx\frac{1}{2\pi\mri}\int_{\lambda-\pi\mri}^{\lambda+\pi\mri}\mrd\lambda'\,
\,\mre^{-\lambda'n+t\alpha_0(\lambda')} Q(\lambda',y,y_0)
\end{align}
with $\lambda>0$.
A saddle point evaluation \cite{bender} of this integral through $\lambda$ obeying
\begin{equation}
\alpha'_0(\lambda)=n/t\equiv u
\end{equation} yields
\begin{equation}
p(n,y,t|y_0)\approx \mre^{-th(n/t)}Q(\lambda(n/t),y,y_0)/\sqrt{2\pi t\alpha_0''(\lambda(n/t))}
\label{eq:pLaplace}
\end{equation} where $'$ denotes a derivative with respect to $\lambda$ from now on.
The exponential term corresponds to the familiar large deviation form with
the rate function given by the Legendre transformation of $\alpha_0(\lambda)$
according to
\begin{equation}
h(u)\equiv u\lambda(u) -\alpha_0(\lambda(u))
\end{equation} which implies
\begin{equation}
\partial_u h(u)=\lambda(u) .
\end{equation} 

The fine structure beyond the large deviation term arises from the
prefactor $Q(\lambda,y,y_0)$ in (\ref{eq:pLaplace}),
which is the only contribution dependent on the hidden variable $y$. We first focus on $n$-values of order 1 which for large $t$ occur with
an exponentially small probability. Still, these values will be relevant for the
fluctuation theorem. Expanding the exponent in (\ref{eq:pLaplace}) to first order in $n/t$ and
then setting $n/t\approx 0$ everywhere leads to
\begin{align}
p(n,y,t|y_0)&\approx \mre^{-th(0) - \lambda_0n}Q(\lambda_0,y,y_0)/\sqrt{2\pi t\alpha_0''(\lambda_0)}\nonumber \\
&= f(t,0)\mre^{-\lambda_0n}Q(\lambda_0,y,y_0)
\label{eq:pn-asy}
\end{align}
where $\lambda_0\equiv \lambda(0)$ is the $\lambda$-value solving $\alpha'_0(\lambda)=0$. The time-dependence is abbreviated by
\begin{equation}
f(t,u)\equiv  \mre^{-th(u)} /\sqrt{2\pi t\alpha_0''(\lambda(u))} .
\end{equation}
The value $\lambda_0$ can be determined analytically since
the convex 
$\alpha_0(\lambda)$ obeys, as will be shown in the general framework below,
the Gallavotti-Cohen symmetry \cite{lebo99,laco08,laco09}
\begin{equation}
  \alpha_0(\lambda)=\alpha_0(-\lambda-(\varDelta \mu -\fe d)),
\end{equation}
which implies 
\begin{equation}
  \lambda_0=-(\varDelta \mu -\fe d)/2.  
\end{equation}
Note that $-2\lambda_0$ corresponds to the entropy production
in the medium associated with a
displacement of motor and bead along one periodicity interval since $\varDelta \mu$ is
the input of chemical work and $\fe d$ is the delivered mechanical work. 
This entropy production is positive in the regime of interest $\fe d<\varDelta\mu$.

For the experiments addressing the FT the relevant quantity is the probability
distribution $p(\dx)$ for the distance
\begin{equation}
\dx\equiv n(t)+[y_0-y(t)]/d
\end{equation} traveled by the bead during time $t$ and measured in units of $d$. This distribution is obtained
by summing $p(n,y,t|y_0)$ over all $n$ and integrating over all initial $y_0$ with the stationary
distribution $\pstat(y_0)$ as
\begin{align}
p(\dx,t)=\int_{-\infty}^\infty&\mrd y_0\,\pstat(y_0)\times\nonumber\\
&\sum_n p(n,y=(n-\dx)d+y_0,t|y_0).
\end{align}
For $\varDelta x=O(1)$ we can plug in (\ref{eq:pn-asy}) and then, for convenience when expressing ratios later on, suppress the overall
time dependence by focusing on
\begin{align}
  &\mathcal{P}(\dx)\equiv\lim_{t\to\infty}p(\dx,t)/f(t,0)\label{eq:motor_u0}\\
&=\int_{-\infty}^\infty\mrd y_0\,\pstat(y_0)\sum_n \mre^{-\lambda_0n}
  Q(\lambda_0,(n-\dx)d+y_0,y_0).\nonumber
\end{align}
While an explicit evaluation requires numerics as presented below, the crucial symmetry 
\begin{equation}
\mathcal{P}(\dx+m)=\mre^ {-\lambda_0m}    \mathcal{P}(\dx)
\end{equation}
for integer $m$ is easily proven by shifting the summation index $n$.

\begin{figure}
\onefigure{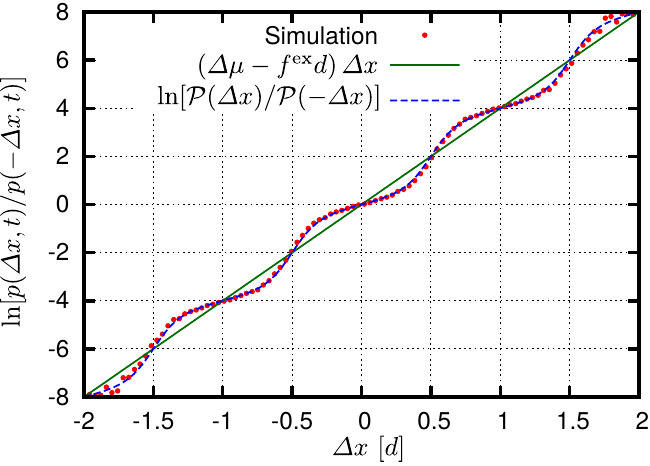}
\caption{
FT-type representation of the long-time asymptotics of the distribution
function $\ln[\mathcal{P}(\dx)/\mathcal{P}(-\dx)]$ (blue dashed line) compared
to $\ln[p(\dx,t)/p(-\dx,t)]$ from simulations of
\eqref{eq:langevin} for $t=1\,\mathrm{sec}$. The straight green line shows the slope corresponding to
entropy production. Parameters: $\gamma=0.41\,\mathrm{sec}/d^2$,
$\varDelta\mu=19$, $\fe=15/d$, $V(y)=(25/2)(y/d)^2$,
$w^+(y)=1\,575\,\mre^{-2.5y/d}\,\mathrm{sec}^{-1}$, $w^-(y)$ determined from (\ref{eq:db_motor}).}
\label{fig:gfunc}
\end{figure}
This discrete symmetry of the probability distribution for the displacement of the
bead is our first main result. It is valid in the long-time limit for all $|\dx|$ of order 1, i.e., for values
of $|\dx|$ that do not scale with time $t$. It implies
that $\mathcal{P}(\dx) = \mre^{-\lambda_0\dx}\phi(\dx)$
with a periodic function $\phi(\dx+1)=\phi(\dx)$, and, in particular $\phi(-1/2)=\phi(1/2)$.
The latter form leads to the FT-type representation
\begin{equation}
\lim_{t\to \infty} \ln\frac{p(\dx,t)}{p(-\dx,t)}=\ln \frac{\mathcal{P}(\dx)}{\mathcal{P}(-\dx)}=- 2 \lambda_0 \dx + \psi(\dx)
\label{eq:ftpsi}
\end{equation}
with a periodic antisymmetric function $\psi(\dx)\equiv \ln [\phi(\dx)/\phi(-\dx)]$ that
vanishes if $\dx$ is evaluated at integer or half integer positions. Hence, the
``slope'' in such an FT-like representation of the distance traveled by
the bead is not constant but rather
periodically modulated around the value corresponding to entropy production.
 In particular, the slope
at $\dx=0$ is {\sl not} given by the entropy production. However, the slope
evaluated as a finite difference over an integer interval chosen symmetrically
around $\dx=0$ allows to infer  the correct entropy production.
In Fig.~\ref{fig:gfunc}, we illustrate this result by comparing simulations of this model with
an explicit calculation of these fine-structured large deviations via a
straightforward numerical evaluation of the largest eigenvalue $\alpha_0(\lambda_0)$, the
corresponding eigenfunction $q_0(\lambda_0,y)$, the expansion coefficient $c_0(\lambda_0,y_0)$
and the stationary distribution $\pstat(y_0)$ for realistic parameters as derived
in \cite{zimm12}. Even for the small value $t=1\,\mathrm{sec}$, the FT representation of the probability distribution
$p(\dx,t)$ is already
in excellent agreement with the numerically calculated $\mathcal{P}(\dx)$ for
the long-time limit. As expected, intersections with the
straight line corresponding to entropy production occur at integer and
half-integer values of $\dx$.

\begin{figure*}
\begin{center}
\includegraphics[scale=0.76]{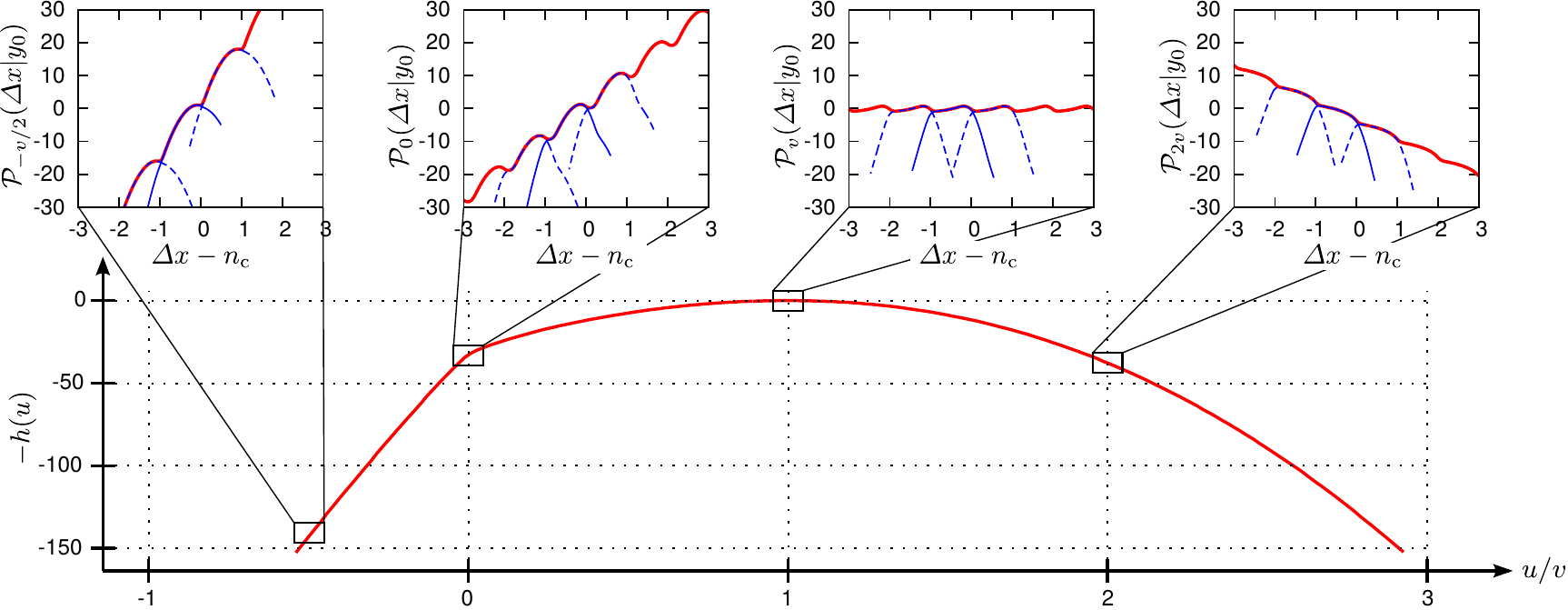}
\caption{Illustration of the properties of the long-time limit of the
  distribution $p_u(n,y,t|y_0)$. The large panel shows the exponential contributions through the rate function $-h(u)$. The small panels show
  the sub-exponential contributions $\mathcal{P}_u(\dx|y_0)$ for selected values of
  $u$ and $y_0=0$ (red curves). The blue curves show the eigenfunctions in the form
  $\mre^{-\lambda_um}Q(\lambda_u,(n_\mathrm{c}+m-\dx)d,0)$ for $m=0$ (solid) and $m=\pm1$
  (dashed). Parameters: $\varDelta\mu=19$, $\fe=0$, $V(y)=20(y/d)^2$, $\gamma=1\,\mathrm{sec}/d^2$,
   $w^+(y)=595\,\mre^{-4y/d}\,\mathrm{sec}^{-1}$.}
\label{fig:zoom}
\end{center}
\end{figure*}
A fine structure as presented above is not confined to a region around $n\simeq \dx\simeq
O(1)$. We can choose any coarse scale $n_\mathrm{c}=ut$. The choice ${u} =0 $ is what we have
just discussed whereas ${u} =v$ with $v$ the mean velocity of the motor (or bead)
 would correspond to the maximum of the distribution.
As sketched in Fig.~\ref{fig:zoom}, we  can then zoom in on a finer scale of order 1 around 
$n_\mathrm{c}={u} t$ to get
\begin{equation}
p_{u}(n,y,t|y_0) \approx f(t,{u})\,\mre^{-\lambda_{u}(n-n_\mathrm{c})}\,Q(\lambda_{u},y,y_0)
\end{equation}
where $\lambda_{u}$ solves $\alpha'_0(\lambda_{u})={u}$. The index ${u}$ at $p_{u}$ indicates
that this result for the fine structure of the distribution is valid for values of $n$ 
that deviate $O(1)$ from the growing
coarse scale $n_\mathrm{c}={u}t$. In the long time limit, we obtain
\begin{align}
  \mathcal{P}_{u}(\dx|y_0)&\equiv\lim_{t\to\infty}\sum_np_u(n,(n-\dx)d+y_0,t|y_0)/f(u,t)\nonumber\\
  &=\sum_n \mre^{-\lambda_{u}(n-n_\mathrm{c})}Q(\lambda_{u},(n-\dx)d+y_0,y_0)
\end{align}
similarly to (\ref{eq:motor_u0}), as
shown in the small panels of Fig.~\ref{fig:zoom}.
As above, the discrete symmetry
\begin{equation}
\mathcal{P}_u(\dx+m|y_0)=\mre^ {-\lambda_{u} m}    \mathcal{P}_u(\dx|y_0) 
\end{equation}
holds for both $m$ and $|\dx-{u}t|$ of order 1.

\begin{figure}
\centerline{\includegraphics[width=\columnwidth]{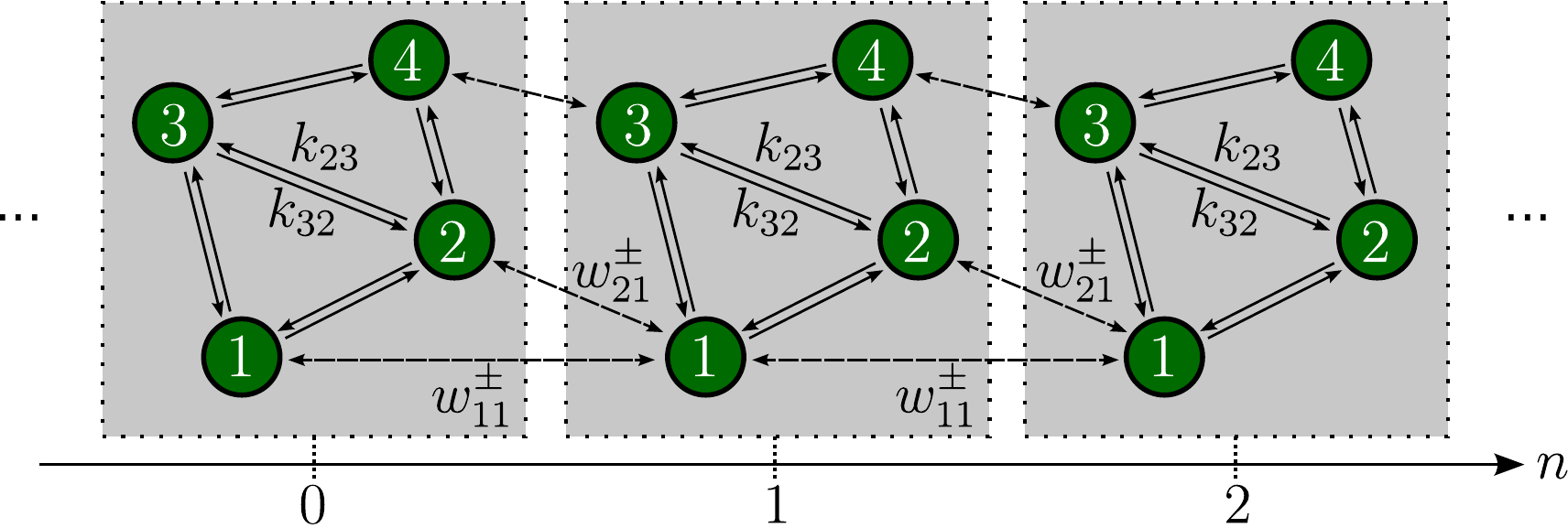}}
\caption{Generalized network. Each square box is a mesostate containing a set of
  microstates (green circles). The mesostates repeat periodically and are
  labeled by the index $n$. Transition rates $k_{ij}$ apply within one
  mesostate and the rates $w_{ij}^\pm$ refer to transitions between
  adjacent mesostates.}
\label{fig:mesostates}
\end{figure}

This approach can easily be generalized to any model with spatial periodicity,
see Fig.~\ref{fig:mesostates}. The states are grouped into identical mesostates labeled by $n$. 
Each mesostate consists of the same  set of microstates labeled by $i$. The state
of the system is thus given by the pair $(n,i)$ which occurs with probability
$p(n,i,t)$. Transitions
between the microstates are possible only if both microstates belong either to
the same mesostate or to an adjacent one. For the first case, the rate from $(n,i)$
to $(n,j)$ is given by $k_{ij}$. The transition rates from
$(n,i)$ to $(n+1,j)$ are given by $w^+_{ij}$ and those from $(n+1,j)$ to $(n,i)$ are given by $w^-_{ij}$.
The spatial periodicity implies that none of the rates depends on $n$. 
A bead attached to a motor is covered by this more general set-up if its continuous position
$y$ is discretized. 
Moreover, this general model allows for
additional substeps of the motor or even a continuous position of the
motor, if the latter is finely discretized. In these cases, the variable $n$ labels the
periodicity interval in which the motor is located, while the actual
position relative to $n$ is incorporated into the state-space $\{i\}$.

Above rates imply that  the evolution of the generating function
\begin{equation}
g_i(\lambda,t) \equiv \sum_{n} \mre^{\lambda n} p(n,i,t)
\end{equation} is given by
\begin{align}
  \partial_t
  g_i(\lambda,t)&=\sum_{j}\left[k_{ji}+\mre^{\lambda}w^+_{ji}
+\mre^{-\lambda}w^-_{ij}-r_i\delta_{ij}
\right] g_j(\lambda,t) \nonumber
\\
&\equiv\sum_j{\mathcal{L}_{ij}(\lambda)}\,g_j(\lambda)
\end{align}
with the  exit rate $r_i\equiv\sum_{\ell}(w^+_{i\ell}+w^-_{\ell i}+k_{i\ell})$.

The matrix $\mathcal{L}_{ij}(\lambda)$ has eigenvalues 
$\alpha_\nu(\lambda)$, with $\alpha_0(\lambda)$ denoting the largest one, 
and right eigenvectors $q_{\nu i}(\lambda)$ obeying 
\begin{equation}
  \sum_j \mathcal{L}_{ij}(\lambda)\,q_{\nu j}(\lambda)=\alpha_\nu(\lambda)\,q_{\nu i}(\lambda).
\end{equation}
Each eigenvalue $\alpha_\nu$ also comes with a left eigenvector 
$\tilde q_{\nu i}(\lambda)$
\begin{equation}
   \sum_i \tilde q_{\nu
     i}(\lambda)\,\mathcal{L}_{ij}(\lambda)=\alpha_\nu(\lambda)\,\tilde q_{\nu
     j}(\lambda)
\end{equation}
with the inner product
\begin{equation}
\sum_i\tilde q_{\nu i}(\lambda)\,q_{\mu i}(\lambda)=C_\nu(\lambda)\,\delta_{\nu\mu}.
\end{equation}
In the long-time limit, the generating function becomes 
\begin{equation}
  g_i(\lambda,t)\approx
\mre^{\alpha_0(\lambda)t} c_0(\lambda,i_0)\,q_{0i}(\lambda).
\end{equation}
If the system starts at time $t=0$ in $(0,i_0)$, the expansion
coefficient is given by $c_0(\lambda,i_0)=\tilde q_{0 i_0}(\lambda)/C_0(\lambda)$.

The fine structure around $n_\mathrm{c}= u t$ 
follows by repeating 
the same steps as above as
\begin{align}
  p_u(n,i,t|i_0)&\approx \mre^{-th(u)-(n-n_\mathrm{c})\lambda_u}\,\frac{\tilde
    q_{0i_0}(\lambda_{u})\,q_{0i}(\lambda_{u})}{C_0(\lambda_{u})\sqrt{2\pi
      t\alpha_0''(\lambda_{u})}} \nonumber \\
&=  f(t,u)\,\mre^{-\lambda_u(n-n_\mathrm{c})}\,Q(\lambda_u,i,i_0)
  \label{eq:general_propagator}
\end{align}
with
\begin{equation}
  Q(\lambda_{u},i,i_0)\equiv \tilde q_{0i_0}(\lambda_{u})\,q_{0i}(\lambda_{u})/C_0(\lambda_u),
\end{equation} which contains the initial microstate $i_0$. This relation implies that
for fixed initial and final states $i_0,i$, the probability for observing
different $n$'s around the coarse scale $n_\mathrm{c}=ut$ obeys
\begin{equation}
  \mathcal{P}_u(n+m,i|i_0)= \mre^{-\lambda_um}  \mathcal{P}_u(n,i|i_0),
\end{equation} 
where we suppress again the time dependence by writing $\mathcal{P}_u(\cdot)\equiv\lim_{t\to\infty}p_u(\cdot,t)/f(u,t)$.
As the introductory model has shown, in an experiment,
 restricting the average to fixed $i$ and $i_0$
may be unrealistic as the direct observation of $n(t)$ might be. For a general
observable defined as
\begin{equation}
\dx(t)\equiv n(t) + z_i(t)-z_{i_{0}}
\end{equation}
that associates a real variable $z_i$ (like displacement) to each microstate $i$, we get as above 
for the fine-structured distribution around the coarse scale $n_\mathrm{c}=u t$ 
\begin{equation}
  p_u(\dx,t)=\sum
  \pstat(i_0)\,p(n,i,t|i_0)
\end{equation}
where $\pstat(i_0)$ is the stationary distribution. In the long-time limit, we have
\begin{equation}
  \mathcal{P}_u(\dx)= \sum
  \pstat(i_0)\,\mre^{-\lambda_un} Q(\lambda_u,i,i_0) 
\end{equation}
where the sums runs over all $n,i,i_0$ with $\dx(n,i,i_0)=\dx$.
As above, the summation implies the discrete symmetry 
\begin{equation}
  \mathcal{P}_u(\dx+m)=\mre^{-\lambda_u m}\mathcal{P}_u(\dx),
\end{equation}
and, 
for $u=0$,  the modulated FT (\ref{eq:ftpsi}).

The discrete symmetry thus derived for a general spatially periodic model
and the identification of a properly discretized slope with $\lambda_0$ is 
universally valid. The identification of this slope with genuine entropy production,
however, 
requires that any cycle either within a mesostate or involving several mesostates
comes with zero entropy production. If this condition is violated, 
entropy production without concomitant displacement occurs. Formally, this condition
means, first, that the rates within each mesostate must
obey the detailed balance condition 
\begin{equation}
  k_{ij}/k_{ji}=\mre^{\varPhi_i-\varPhi_j}.
  \label{eq:general_db}
\end{equation}
The generalized potential $\varPhi$ comprises not only the
internal energy of the system but also external contributions like
displacement against an external force or the consumption of molecules
associated with a fixed chemical potential \cite{seif11a}. Second, the rates
between adjacent mesostates must fulfill 
\begin{equation}
  w^+_{ij}/w^-_{ij}=\mre^{\varPhi_i-\varPhi_j+\varDelta S}
  \label{eq:general_ldb}
\end{equation}
with $\varDelta S$ independent of $i$ and $j$. Straightforward algebra then shows
\begin{equation}
  \mre^{-\varPhi_i}\mathcal{L}_{ji}(\lambda)\,\mre^{\varPhi_j}=
  \mathcal{L}_{ij}(-\lambda-\varDelta
  {S}).
  \label{eq:db_trafo}
\end{equation}
Hence,  $\mathcal{L}(\lambda)$ and
$\mathcal{L}(-\lambda-\varDelta S)$ have the same spectrum
which implies the 
Gallavotti-Cohen symmetry
\begin{equation}
  \alpha_0(\lambda)=\alpha_0(-\lambda-\varDelta S)
\label{eq:GC_general}
\end{equation}
and the relation $\mre^{-\varPhi_i}\tilde q_{\nu i}(\lambda)=q_{\nu i}(-\lambda-\varDelta S)$
between the left and right eigenvectors.
From (\ref{eq:GC_general}) the fluctuation theorem-type relation 
 $h(u)=h(-u)-u\varDelta S$
follows for the rate function $h(u)$. In this case, $\varDelta S = -2\lambda_0$ indeed corresponds to
 the
entropy production associated with 
stepping one periodicity unit. This insight can be turned into a
simple experimental check for the presence of idle cycles that burn ATP without
leading to net displacement\cite{yasu98,nish04,rond05,gebh06}. If, in the
long-time limit, the properly discretized slope in the FT representation
of the displacement deviates from $\varDelta \mu-\fe d$, then the motor must have
 idle cycles.

Colloidal particles either driven externally over periodic potentials \cite{spec07,gome09} or
subject to active motion along periodic substrates \cite{volp11} are a second class of
systems for which the present approach could lead to a refined picture. A
particularly intriguing observation in the first case was the occurrence of a
kink around zero entropy production \cite{mehl08,doro11,spec12}, also visible here in
Fig.~\ref{fig:zoom}. It will also be interesting to explore within this
approach the deviations from the ideal FT observed
experimentally in such systems \cite{mehl12}.

In conclusion, we have introduced the concept of fine-structured large deviations that
leads to a periodic modulation of the probability distribution for displacements of stochastic
motion along periodic substrates. 
It explains the deviations from a constant slope in an FT-like representation
of this distribution, which are characteristic for systems with hidden degrees
of freedom.
Applied to molecular motors, our approach reveals how to infer
entropy production from the properly discretized slope provided there 
are no idle cycles. The application
of these concepts  to driven or 
active colloidal motion along a periodic structure will be left for future work.

\acknowledgments
We thank S. Toyabe for valuable interactions.

\section{Appendix}
\begin{figure}
\onefigure{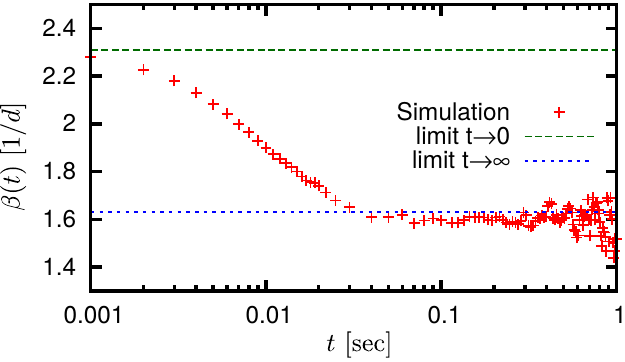}
\caption{
Time dependence of the FT-slope $\beta(t)$ (\ref{eq:betaslope}). While the short-time limit corresponds to the mean force
  exerted on the bead (green line at $\mean{\partial_yV(y)}-\fe=2.31/d$), the long-time limit has no physical interpretation
  beyond being the slope of the FT representation of $\mathcal{P}(\dx)$ at $\dx=0$
  (blue line at $1.63/d$). Parameters as in Fig.~\ref{fig:gfunc}.}
\label{fig:alpha_t}
\end{figure}
The average force that the motor exerts on the 
bead that has tentatively been identified with the long-time limit value of the
slope 
\begin{equation}
  \label{eq:betaslope}
  \beta(t)\equiv\lim_{\dx\to 0}\frac{1}{\dx\,d}\ln\frac{p(\dx,t)}{p(-\dx,t)}
\end{equation}
at $\dx=0$ in \cite{haya10,haya12} is rather given by the short-time limit of this slope as we
now show. Consider a bead subject to an external force $\fe$ and coupled to a motor in a steady state. The total force
$F$ exerted on the bead is a random variable with an underlying distribution
$\pstat(F)$. For short times $t$, the distribution of displacements 
of the bead corresponds to a diffusion biased by this force. Taking the average over $F$ yields
\begin{equation}
p(\dx,t)\approx \int_{-\infty}^\infty \mrd F\,\frac{\pstat(F)}{d\,\sqrt{4\pi t/\gamma}}\,
{\mre^{-(\dx d-Ft/\gamma)^2\gamma/(4t)}}.
\end{equation} 
Simple calculus then shows that
\begin{equation}
\lim_{t\to 0}\beta(t)=\int_{-\infty}^\infty \mrd F\,\pstat(F)\,F=\mean{F}=-\fe+F^\mathrm{m},
\end{equation}
where $F^\mathrm{m}$ is the average force the motor exerts via the linker on
the bead.
In the case of the simple model presented in the main text, it is given by
$F^\mathrm{m}=\mean{\partial_yV(y)}$. Simulations
of the full model (see Fig.~\ref{fig:alpha_t}) confirm the validity of this simple derivation.

\bibliographystyle{eplbib_doi}
\bibliography{/home/public/papers-softbio/bibtex/refs}

\end{document}